\documentclass{elsart}
\usepackage{ifpdf}
\usepackage{graphicx,natbib,amssymb,lineno}
\usepackage{graphics}
\usepackage[%
  breaklinks=true,%
  colorlinks=true,%
  linkcolor=blue,anchorcolor=blue,%
  citecolor=blue,filecolor=blue,%
  menucolor=blue,pagecolor=blue,%
  urlcolor=blue]{hyperref}

\makeatletter
\def\elsartstyle{%
    \def\normalsize{\@setfontsize\normalsize\@xiipt{14.5}}
    \def\small{\@setfontsize\small\@xipt{13.6}}
    \let\footnotesize=\small
    \def\large{\@setfontsize\large\@xivpt{18}}
    \def\Large{\@setfontsize\Large\@xviipt{22}}
    \skip\@mpfootins = 18\p@ \@plus 2\p@
    \normalsize
}
\@ifundefined{square}{}{}
\makeatother

\begin{document}

\begin{frontmatter}
\title{On the origin of the anomalous behaviour of 2$^+$ excitation energies
in the neutron-rich Cd isotopes}
\author[UAM]{Tom\'as R. Rodr\'iguez}, 
\ead{tomas.rodriguez@uam.es}
\author[UAM]{J. Luis Egido\corauthref{cor}},
\corauth[cor]{Corresponding author.}
\ead{j.luis.egido@uam.es}
\author[CSIC]{ Andrea Jungclaus} 
\ead{andrea.jungclaus@iem.cfmac.csic.es}
\address[UAM]{Departamento de F\'isica Te\'orica C-XI, Universidad Aut\'onoma de Madrid, 28049 Madrid, Spain}
\address[CSIC]{ Instituto de Estructura de la Materia, Consejo Superior de Investigaciones Cient\'ificas, 28006 Madrid, Spain}

\begin{abstract}
Recent experimental results obtained using $\beta$ decay and isomer spectroscopy indicate an unusual behaviour of the energies of the first excited 2$^{+}$ states in neutron-rich Cd isotopes approaching the N=82 shell closure. To explain
the unexpected trend, changes of the nuclear structure far-off stability have been suggested, namely a quenching
of the N=82 shell gap already in $^{130}$Cd, only two proton holes away from doubly magic $^{132}$Sn. We study the behaviour of the 2$^+$ energies in the Cd isotopes from N=50 to N=82, i.e. across the entire span of a major
neutron shell using modern beyond mean field techniques and the Gogny force. We demonstrate that the observed low 2$^+$
excitation energy in $^{128}$Cd close to the N=82 shell closure is a consequence of the doubly magic character of this nucleus for oblate deformation favoring thereby prolate configurations rather than spherical ones.
\end{abstract}
\end{frontmatter}


\begin{figure}[b!]
{\centering {\includegraphics[angle=0,width=1.0\columnwidth]{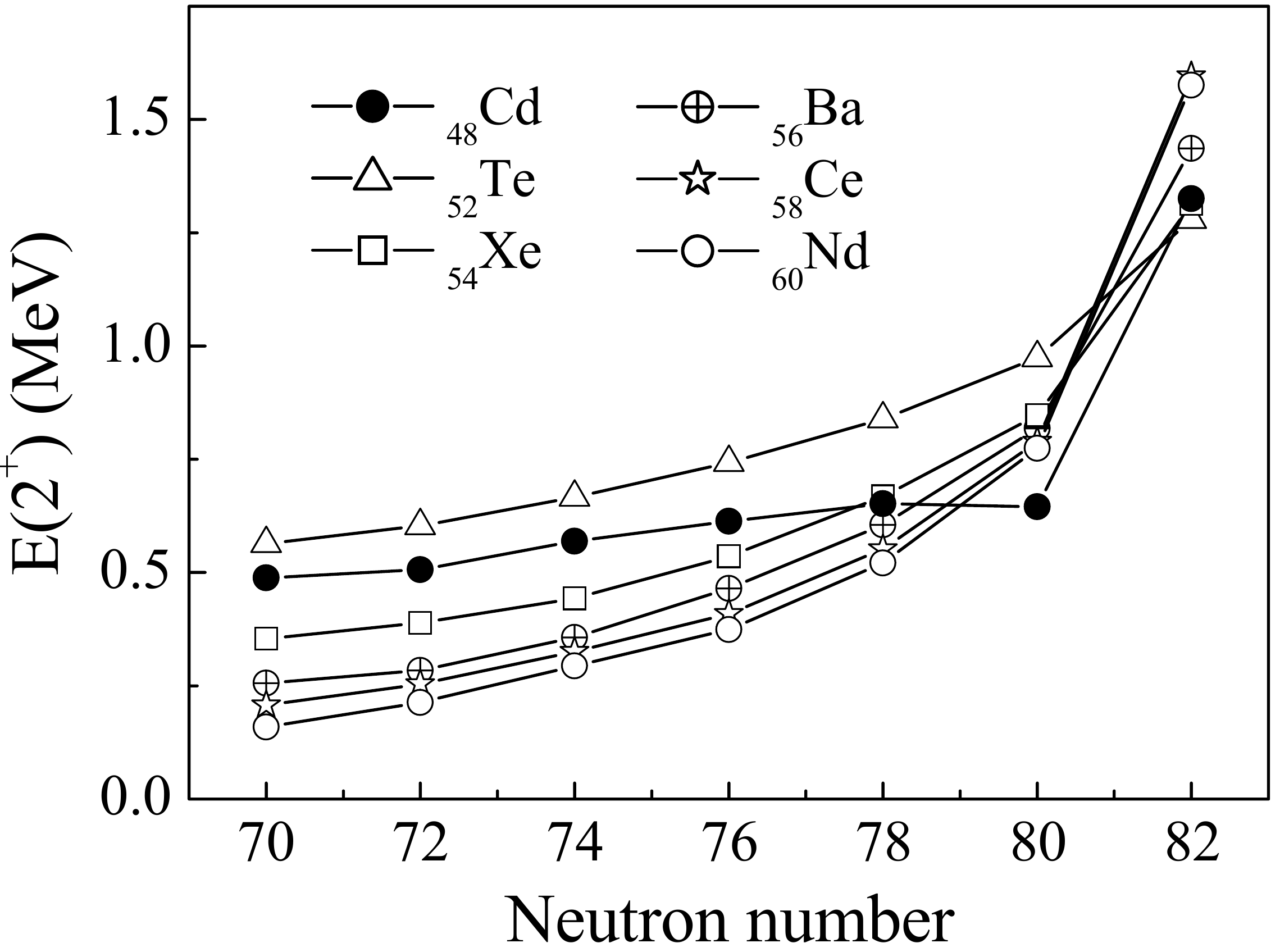}} \par}
\caption{Excitation energies of the first excited 2$^+$ states as a function of the neutron number N
for even-even Cd, Te, Xe, Ba, Ce and Nd isotopes in the range  N=70-82 shell.}
\label{Fig:energies}
\end{figure}
Recent advances in experimental techniques allow to follow the isotopic chains of heavy nuclei and therefore the
evolution of nuclear structure over wider and wider isospin ranges, in some cases even across the entire span of a
major neutron shell. The Cd isotopes (Z=48) for example are the first case, where experimental information about
excited states is now available from the N=50 to the N=82 shell closure, that is from $^{98}$Cd to $^{130}$Cd
\cite{gor97,jun07}. Approaching the proton and neutron drip lines new facets of nuclear structure are
expected \cite{DripLineReview} to gain relevance due to the increased importance of the coupling to the unbound continuum and changes in the density distributions of protons and neutrons. The availability of experimental information over a wide isospin range therefore allows for the most stringent test of the validity of global nuclear structure models.
Besides the nuclear structure aspect, the understanding of the nuclei around the doubly-magic
tin isotopes $^{100}$Sn and $^{132}$Sn is also of astrophysical interest. The rapid proton ($rp$) capture process
of hydrogen burning on the surface of an accreting neutron star ends in a closed SnSbTe cycle near $^{100}$Sn
\cite{sch01} and the dynamics of the $r$ process and with it the solar abundance distribution in the mass 130 region 
is determined by the properties of the N=82 waiting point nuclei below $^{132}$Sn \cite{kra93}. Already more than
ten years ago it was shown that the assumption of a quenching of the N=82 neutron shell closure leads to a
considerable improvement in
the global abundance fit in $r$ process calculations \cite{che95,pfe97}. Different theoretical calculations indeed
predict such a reduction of the shell gap near the neutron drip line \cite{dob94,sha02}. Unfortunately, this region
of very neutron-rich waiting point nuclei is still out of reach experimentally. However, in recent years a number
of experimental observations in the $^{132}$Sn region have been interpreted as first experimental evidence of the
quenching of the N=82 shell closure in $^{130}$Cd \cite{kau00,dil03}, much closer to $^{132}$Sn than predicted
by any calculation. One of them concerns the anomalous behaviour of the energies of the first excited 2$^+$ states
in the heavy Cd isotopes as illustrated in Fig.~\ref{Fig:energies}. The 2$^+$ state in the N=80 isotope $^{128}$Cd at
an excitation
energy of 645 keV lies 7 keV lower than the 2$^+$ level in $^{126}$Cd \cite{kau00}. It is evident from Fig.~1 that the
neutron-rich Cd isotopes display a completely different behaviour when approaching N=82 than the Te, Xe, Ba, Ce and Nd isotopes above the Z=50 shell closure. However, the recently observed excitation scheme of $^{130}$Cd and its
comparison to shell model calculations does not give any evidence for a reduced N=82 gap in this nucleus \cite{jun07}. Furthermore, a new direct mass measurement for $^{134}$Sn restored the $^{132}$Sn shell closure to that expected 
for a doubly-magic nucleus \cite{dwo08}. Previous $Q_\beta$ measurements seemed to indicate a reduced strength of
that shell closure. In light of these new experimental results the interpretation of the anomalous behaviour
of the 2$^+$ energies as hint to a possible N=82 shell quenching has to be questioned.

In this Letter we study the evolution of 2$^+$ excitation energies in the Cd isotopes between N=50 and N=82 from a theoretical view point in an attempt to unveil the origin of the anomalous behaviour.
This endeavor of a realistic calculation across an entire major shell can only be realized in the framework of modern beyond mean field theories with effective interactions \cite{BHR.03,rod02,Ni.06}. An exploratory study of the neutron-rich Cd isotopes has already been presented in \cite{jun06}. In that reference, a simplified approach, neglecting some relevant exchange terms of the Gogny force, has been used and only a few isotopes have been investigated. Furthermore, no particle number projection has
been performed, neither at the mean field level nor in the configuration mixing approach. 

In this work we use the recently proposed symmetry conserving configuration mixing approach \cite{rod07} with
particle number and angular momentum projection and the finite range density dependent Gogny interaction 
(D1S parametrization \cite{BERGNPA84}) with all exchange terms  included.
This interaction has been adjusted to reproduce global properties of nuclear matter more than twenty five years
ago and does not invoke any parameter adjustment to certain regions of the nuclear chart.
Its strength is the ability to describe in a consistent way a large variety of phenomena such as the erosion
of the N=20 shell closure around $^{32}$Mg \cite{rod02}, the emergence of the new shell closure at N=32 in
neutron-rich Ca, Ti, and Cr isotopes \cite{rod07}, the shape coexistence in the Pb isotopes \cite{egi04} as
well as the fission barriers in $^{254}$No \cite{egi00}. The intention of these calculations is to describe
general trends and evolutions of nuclear properties and to help to unravel the basic underlying origins rather
than to obtain an as detailed as possible description of a single nucleus of interest.

\begin{figure}[t!]
{\centering {\includegraphics[angle=0,width=1.0\columnwidth]{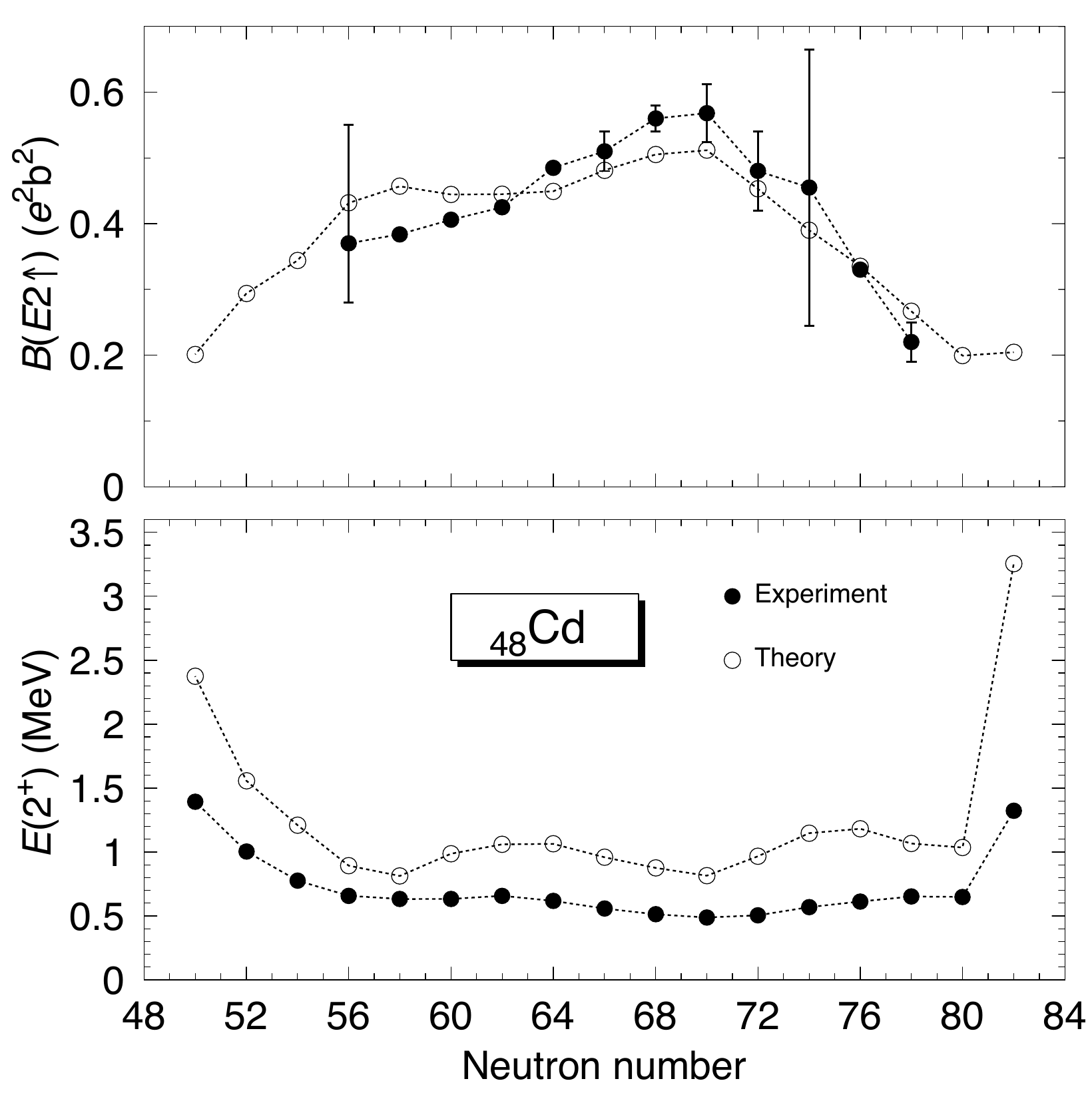}} \par}
\caption{$B(E2)$ transition probabilities (top) and 2$^+$ excitation energies (bottom) for the Cd isotopes
between N=50 and N=82.}
 \label{Fig:E2-BE2}
\end{figure}

 In our approach \cite{rod07}, we first generate the collective subspace of Hartree-Fock-Bogoliubov (HFB) wave functions, $\{|\phi({q})\rangle \}$, by the quadrupole constrained particle number projection (PNP) {\it before the variation} \cite{PNP-Gogny}, i.e., we minimize the energy
\begin{eqnarray}
E^{N,Z}({q})=\frac{\langle{\Phi}^{N,Z}({q})|\hat{H}- \lambda_q \hat{Q}|{\Phi}^{N,Z}({q})\rangle}{\langle{\Phi}^{N,Z}({q})|{\Phi}^{N,Z}({q})\rangle}, 
\label{projener}
\end{eqnarray}
with $|\Phi^{N,Z}({q})\rangle=\hat{P}^{N}\hat{P}^{Z}|\phi({q})\rangle$, 
where $\hat{P}^{N(Z)}$ is the projector onto neutron (proton) number and  $\lambda_q$ the Lagrange multiplier determined by the constraint  $\langle \varphi| \hat{Q}|\varphi\rangle/\langle \varphi|\varphi\rangle=q$, in an obvious notation. The constraint on the operator $ \hat{Q}$   allows the definition of potential energy surfaces along the most relevant degrees of freedom. In the present calculations  we restrict this set to the quadrupole deformation
$\beta$, triaxial shapes, i.e. the $\gamma$ degree of freedom, are not considered. 

The configuration mixing calculation is performed within the generator coordinate method (GCM) framework taking linear combinations of the particle number projected  wave functions obtained in the first step and performing in addition angular momentum projection: 
\begin{eqnarray}
|\Psi^{N,Z}_{J,\sigma}\rangle=\int f^{N,Z}_{J,\sigma}({q}) \; \hat{P}^{J}\;|{\Phi}^{N,Z}({q})\rangle d{q}.
\label{GCMstate}
\end{eqnarray}
with $\hat{P}^{J}$ the angular momentum projector.

Then, the variational principle applied to the weights $ f^{N,Z}_{J,\sigma}({q})$ gives the generalized eigenvalue problem (Hill-Wheeler equation):
\begin{eqnarray}
\int \left(\mathcal{H}^{N,Z}_{J}({q},{q}^\prime)-E^{N,Z}_{J,\sigma}\mathcal{N}^{N,Z}_{J}({q},{q}^\prime )\right)f^{N,Z}_{J,\sigma}({q}^\prime)d{q}^\prime=0
\label{HWeq}
\end{eqnarray}
with  $\mathcal{H}^{N,Z}_{J}$ and $\mathcal{N}^{N,Z}_{J}$ the Hamiltonian and norm overlaps, respectively (see \cite{rod02} for further details). In the calculations we use 11 harmonic oscillator major shells and the  D1S parametrization \cite{BERGNPA84} of the Gogny force.

It is important to emphasize here that nuclei close to shell closures belong to the weak pairing
regime where the HFB approach breaks down. To properly deal with the pairing correlations in
nuclei such as $^{132}$Te and $^{128}$Cd with only a few particles/holes outside a doubly-magic core
it is mandatory to apply the so-called Projection Before Variation approach as described above.  As a matter of fact we find quantitative differences between the present calculations and those of Ref.~\cite{jun06} performed also
with the Gogny interaction but without PNP and without exchange terms.
Furthermore,  in order to check the validity of the axial approach used in this work, we have also performed triaxial calculations at the PNP level. The results show that none of the analyzed isotopes  exhibits   triaxial minima or large $\gamma$  softness \cite{TL.09}.

\begin{figure}[t!]
{\centering {\includegraphics[angle=0,width=1.0\columnwidth]{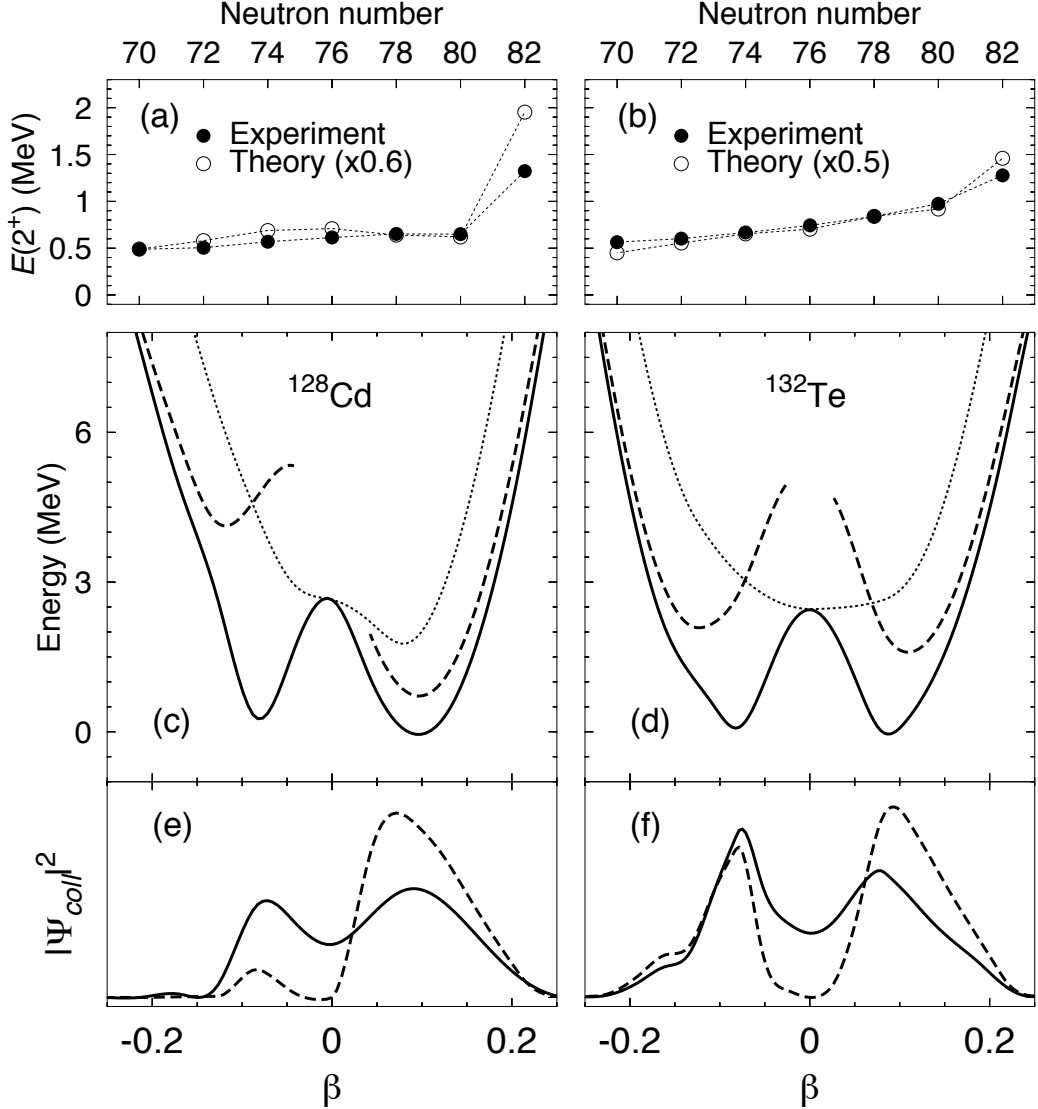}} \par}
\caption{ a),b)  Calculated and experimental 2$^+_1$ energies in the Cd and Te isotopes as a function of the neutron number; 
c),d) potential energy curves as a function of the intrinsic quadrupole deformation $\beta$, with particle number projection (dotted  lines) and after additional angular momentum projection, $J=0 \hbar$ (continuous lines) and $J=2 \hbar$  (dashed lines); e),f) squared amplitudes of the wave functions of the collective states of Eq.~\ref{GCMstate}  with $J=0 \hbar$ (continuous lines) and $J=2 \hbar$ (dashed lines).}
 \label{Fig:bmf-cd-te}
\end{figure}

In Fig.~\ref{Fig:E2-BE2}  we compare the calculated excitation energies of the  $2^+_1$ states (the eigenvalues of Eq.~\ref{HWeq} for $J=2^+$ and $\sigma=1$) and the transition strengths $B(E2, 0^+\rightarrow 2^+)$ for all Cadmium isotopes from the N=50 to the N=82 shell closure to the available experimental information \cite{NUCDATA}. We find an overall good qualitative agreement between theory and experiment. The general trend as well as the absolute strength of the experimental $B(E2)$ values are well reproduced by our calculations.  For the $2^+$  excitation energies one can distinguish   three  different regions. In the first one, N=51-58, the parabolic behaviour of the energy roughly  corresponds to the successive filling of the g$_{7/2}$ neutron orbital.  In the second region, N=59-70, with a rather flat energy, several smaller sub-shells (d$_{5/2}$,  d$_{5/2}$, s$_{1/2}$) are being filled. Finally, from N=71 to 82, the h$_{11/2}$ orbital is being filled, but at variance from the first region here we observe the unexpected non-parabolic behaviour of the   $2^+$ energies. These three regions have an approximate counterpart in the transition probabilities.
 Though we nicely reproduce these features  the calculated energies are too high on an absolute scale.  The origin of this scaling is supposedly due to the absence of K-mixing in the calculations. This mixing can be induced in different, non-independent, ways:  at the HFB level, by angular momentum projection before variation \cite{carlo}, by the Coriolis force (cranking) or by admixing of two-quasiparticle states (Tamm-Dancoff) \cite{carlo,sun}, and at the configuration mixing level by the consideration of triaxial shapes \cite{BH-08}.  Within our approach we have investigated the $(\beta,\gamma)$ plane  and, as mentioned above, none of the nuclei studied show triaxial minima or large $\gamma$ softness and it seems that in this case the consideration of K-mixing effects amounts, at least for the lowest lying collective states,  to a global, rather constant, change in the  mass parameter (either for rotations or vibrations). That means, the ratio between the calculated and experimental $2^+$ energies is nearly constant (about 1.6) over a large range of isotopes. This effect, which is not yet fully understood, has already been observed in previous studies, see for example \cite{rod07}.  A rather constant ratio has also been obtained in limiting situations evaluating the mass parameter with and without triaxial effects \cite{ER.04} and in an exact triaxial angular momentum projection \cite{doba}.
Taking into account this constant factor the calculations reproduce the experimental data even quantitatively. This
is shown in Fig.~\ref{Fig:bmf-cd-te}(a,b) where the renormalized calculated 2$^+$ excitation energies for the Cd and Te isotopes are compared  to experimental data in the region of interest close to N=82. The calculations nicely reproduce the experimental trends, namely the parabola-like behaviour in the Te isotopes and the ``pathological'' flattening up to N=80 for the Cd isotopes. 

To trace back the
origin for the different behaviour of the N=76, 78, and 80 Cd and Te isotones we will compare in more detail the different steps of the calculations for the N=80 pair. In Fig.~\ref{Fig:bmf-cd-te}(c,d) we present the potential energy curves calculated {\it before} configuration mixing as a function of the intrinsic quadrupole deformation $\beta$. These curves
show clear differences between $^{128}$Cd and $^{132}$Te and shed light on the origin of the different
behaviour of the 2$^+$ energies. The energies obtained using the solutions of the PNP equations (see eq.~\ref{projener}) show an intrinsic spherical minimum in $^{132}$Te and a slightly prolate one in $^{128}$Cd. 
The results obtained after angular momentum projection, i.e., the energy expectation values calculated with the wave functions $|\Phi^{N,Z}_J({q})\rangle=\hat{P}^{J}\hat{P}^{N}\hat{P}^{Z}|\phi({q})\rangle$, are rather similar for both nuclei for the 0$^+$ 
state: They present coexisting prolate and oblate minima. However, the results for the 2$^+$ state are rather
different: for $^{128}$Cd we obtain a deep prolate deformed minimum about 3.5 MeV lower than the oblate one,
whereas in $^{132}$Te we observe again two symmetric minima as for the 0$^+$ state. Finally, in
 Fig.~\ref{Fig:bmf-cd-te}(e,f) the wave functions of the collective states (eq.~\ref{GCMstate}) with angular
momentum 0 and 2, respectively, i.e. the states obtained {\it after} configuration mixing, are shown (the corresponding 2$^+$ energies have already been shown in Fig.~\ref{Fig:bmf-cd-te}(a,b)). Once more, pronounced
differences are observed for the 2$^+$ state: In $^{132}$Te this state represents on average a quadrupole
oscillation around a spherical shape. In $^{128}$Cd, on the other hand, the 2$^+$ state is essentially an
oscillation around a prolate deformed shape and therefore strongly coupled to rotation. This is the main reason
why the 2$^+$ energies in the heavy Cd isotopes behave differently from those in the heavy Te isotopes.
\begin{figure}[t!]
{\centering {\includegraphics[angle=0,width=1.0\columnwidth]{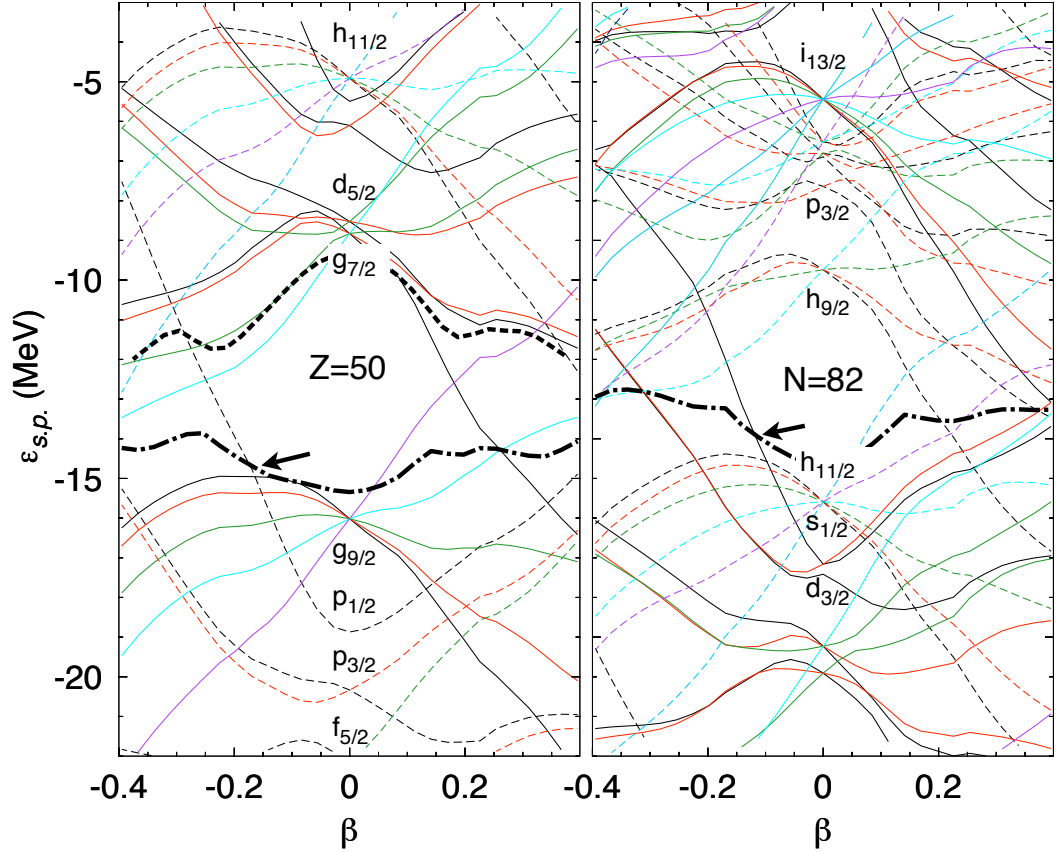}} \par}
\caption{Proton (left) and neutron (right) single particle energies for $^{128}$Cd. The energy scale for neutrons has been shifted by 9 MeV. The thick dotted-dashed lines represent  the Fermi levels for $^{128}$Cd and the thick dashed line the proton Fermi level for $^{132}$Te.}\label{Fig:sp-energies}
\end{figure}
Trying to understand the underlying origin of the differences between the properties of the 2$^+$ states in
$^{128}$Cd and $^{132}$Te
we will now have a closer look to the potential energy curves obtained after particle number but before angular
momentum projection (dotted lines in Fig.\ref{Fig:bmf-cd-te}(c,d)). Already here, a difference between the two
nuclei is apparent. On the oblate side ($\beta < 0$), the energy increases much faster for $^{128}$Cd compared
to $^{132}$Te ($E(\beta=-0.1)-E(\beta=0)$=2.5 MeV and 1.0 MeV, respectively). This difference can be understood
on the basis of the single particle energies (SPE), represented in Fig.~\ref{Fig:sp-energies} for
$^{128}$Cd.  The SPE  for $^{132}$Te
are not significantly different from these ones. The proton Fermi levels for both nuclei are included in
Fig.~\ref{Fig:sp-energies}, while for the neutrons we assume the Fermi level to be the same since both
nuclei have the same number of neutrons, $N=80$. At an oblate deformation of about $\beta=-0.1$ the neutron
$1s_{1/2}$ level is emerging from the Fermi surface, see the corresponding arrow. From here on both N=80 isotones have a completely filled
$\nu h_{11/2}$ shell. On the proton side, a similar situation occurs around $\beta=-0.15$ where the proton
$1p_{1/2}$ level is crossing the Fermi surface for $^{128}$Cd. That means that for oblate
deformations larger than $|\beta| \approx 0.15$ in addition to the full $h_{11/2}$ neutron shell the nucleus
$^{128}$Cd also has a completely filled large intruder proton orbit, namely the $\pi g_{9/2}$ shell.
To deform this nucleus with closed large intruder shells for both protons and neutrons a lot of
energy is required resulting in the very steep PNP curve for $^{128}$Cd in Fig.~\ref{Fig:bmf-cd-te}(c).
Furthermore, the generation of two units of angular momentum to form a 2$^+$ state is very costly for such
a stiff configuration since it requires the admixing of orbitals above the $Z=50$ or $N=82$ shell closures
(compare the curve for $J=2^+$ curve in Fig.~\ref{Fig:bmf-cd-te}(c)). The situation in $^{132}$Te is very
different. In this case, the proton Fermi level lies above the $Z=50$ shell closure and the oblate branch of the proton system
can easily be deformed (see Fig.~\ref{Fig:sp-energies}) due to the positive slope of the relevant $g_{7/2}$
orbitals. This deformation will be directly transfered to
the neutron system due to the strong proton-neutron interaction.
Even more important, a $J=2^+$ state can easily be build by coupling two $g_{7/2}$ protons (see Fig.~\ref{Fig:bmf-cd-te}(d)). To summarize this discussion, we have shown that the prolate deformation of the
2$^+$ state in $^{128}$Cd has its origin in a very peculiar coincidence of closed shell proton and neutron
configurations for $\beta < -0.15$ leading to a ''blocking'' of the oblate branch of the energy curve, i.e., in this branch
the nucleus behaves as a double magic one.  Since the  proton oblate blocking persist also   in
the nearby $N=76$ and 78 Cadmium isotopes it is obvious that these isotopes will be on average more deformed than the 
corresponding Tellurium isotopes, as depicted by the $2^+$ energies in Fig.\ref{Fig:bmf-cd-te}(a,b), indicating  that the different behavior of the Cadmium and Tellurium isotopes close to the $N=82$ shell closure can be explained by "standard" nuclear structure effects without invoking shell quenching.

In conclusion we have presented a theoretical study of the origin of the unusual behaviour of 2$^+$ excitation energies
in neutron-rich Cadmium isotopes towards the N=82 shell closure using modern beyond mean field techniques and the Gogny
force. We have demonstrated that these calculations, which do not dispose of any adjustable parameters, are not
only able to describe general features of atomic nuclei across entire major shells, i.e. over wide ranges
of isospin, but also to reproduce local features of nuclear structure such as the low 
2$^+$ energies in $^{126,128}$Cd and unravel their origin.
Furthermore, our calculations show that the anomalous behaviour of the 2$^+$ energies in the heavy Cd isotopes 
is caused by the very special characteristics  of  the Cd isotopes which favor prolate configurations close to the $N=82$ shell closure. There is no need to assume any changes of nuclear structure far-off stability, i.e. a quenching of the $N=82$ shell closure, to explain the experimental findings.

The authors acknowledge financial support from the Spanish Ministerio de Educaci\'on y Ciencia 
under contracts FPA2005-00696, FPA2007-66069 , by the Spanish Consolider-Ingenio 2010
Programme CPAN (CSD2007-00042) and within the programa Ram\'on y Cajal (A. Jungclaus). 



\end{document}